\documentclass{aastex}
\usepackage{emulateapj5}
\usepackage{apjfonts}

\submitted{Accepted by the Astrophysical Journal 2003 July 7}

\shortauthors{Wilson et al.}
\shorttitle{High-Resolution Continuum Imaging of W3 IRS 5}

\begin{document}

\title{High-Resolution Continuum Imaging at 1.3 and 0.7~cm of the W3 IRS 5 
Region} 
\author{T. L. Wilson\altaffilmark{1}, 
D. A. Boboltz\altaffilmark{2}, 
R. A. Gaume\altaffilmark{2},
S. T. Megeath\altaffilmark{3}}

\altaffiltext{1}{Max-Planck-Institut f\"ur Radioastronomie, 
Postfach 2024, D-53010 Bonn, Germany}
\altaffiltext{2}{U. S. Naval Observatory,
3450 Massachusetts Ave., NW, Washington, DC 20392-5420}
\altaffiltext{3}{Harvard-Smithsonian Center for Astrophysics, 60 
Garden Street, Cambridge, MA 02138}

\begin{abstract}  

High-resolution images of the hypercompact HII regions (HCHII) in W3
IRS 5 taken with the Very Large Array (VLA) at 1.3 and 0.7~cm are
presented.  Four HCHII regions were detected with sufficient
signal-to-noise ratios to allow the determination of relevant
parameters such as source position, size and flux density.  The
sources are slightly extended in our $\sim$0.2$''$ beams; the
deconvolved radii are less than 240 AU.  A comparison of our data with
VLA images taken at epoch 1989.1 shows proper motions for sources IRS
5a and IRS 5f.  Between 1989.1 and 2002.5, we find a proper motion of
210~mas at a position angle of 12$^{\circ}$ for IRS 5f and a proper
motion of 190~mas at a position angle of 50$^{\circ}$ for IRS 5a.  At
the assumed distance to W3 IRS 5, 1.83$\pm$0.14 kpc, these offsets
translate to proper motions of $\sim$135~km~s$^{-1}$ and
$\sim$122~km~s$^{-1}$ respectively. These sources are either shock
ionized gas in an outflow or ionized gas ejected from high mass
stars. We find no change in the positions of IRS 5d1/d2 and IRS 5b;
and we show through a comparison with archival NICMOS 2.2 $\mu$m
images that these two radio sources coincide with the infrared double
constituting W3 IRS 5.  These sources contain B or perhaps O stars.
The flux densities of the four sources have changed compared to the
epoch 1989.1 results.  In our epoch 2002.5 data, {\it none} of the
spectral indicies obtained from flux densities at 1.3 and 0.7~cm are
consistent with optically thin free-free emission; IRS 5d1/d2 shows
the largest increase in flux density from 1.3 cm to 0.7 cm.  This may
be an indication of free-free optical depth within an ionized wind, a
photoevaporating disk, or an accretion flow. It is less likely that
this increase is caused by dust emission at 0.7 cm.
\end{abstract}

\keywords{H II regions--ISM:individual(W3)--radio continuum:ISM--stars: formation}

\section{INTRODUCTION}

W3 is a high-mass star formation region located in the Perseus arm at
a distance of 1.83$\pm$0.14 kpc as determined from H$_2$O maser proper
motions \citep{IMAI:00}.  W3 contains several sites of star formation,
the most active of which is W3 Main, which contains at least 10 young
high mass stars and protostars within a region spanning only a few
parsecs.  The molecular gas in W3 Main is distributed into two
distinct molecular clumps \citep{TWSGJC:95}.  The eastern clump
appears to be the more active of the two clumps and is associated with
two bright, extended HII regions. The more evolved, shell-like HII
region, W3A, lies to the east of the more compact HII region W3B
\citep[see][hereafter TGC97]{TGCWJ:97}.

Located between W3A and W3B is the luminous infrared source W3 IRS 5.
This source was first discovered by \cite{WBN:72}.  Because of the
lack of radio continuum emission toward IRS 5, \cite{WBN:72} suggested
that IRS 5 was in a protostellar phase.  Subsequent observations at
infrared and submillimeter wavelengths showed a steeply rising
spectral energy distribution from 2 to 50 $\mu$m with a peak at 100
$\mu$m and a total luminosity greater than $10^5$ L$_\odot$
\citep{CBHECS:95}.  Infrared slit-scan and speckle measurements
resolved IRS 5 into a double source with a separation of 1$''$
\citep{NBM:82, HML:81}.  Wide-field infrared imaging has detected a
dense cluster of at least 80 low mass stars surrounding W3 IRS 5;
these stars are distributed over a region 40$''$ (0.5 pc) in diameter
\citep[see][]{MHBGHRS:96}. Chandra observations have detected a X-ray
source toward W3 IRS 5, making IRS 5 a rare system of massive young
stellar objects that can be studied from X-ray to radio wavelengths
\citep{HDWCL:02}.

Molecular line observations have produced ample evidence that IRS 5 is
the source of at least one molecular outflow. CO observations show a
bipolar outflow centered on IRS 5 with an axis oriented in
northeast--southwest direction and a position angle of $\sim$38$^{\rm
  o}$ \citep{CLAUSSEN:84,MMH:91}. Proper motion measurements of a
cluster of H$_2$O masers surrounding W3 IRS 5 show an overall motion
away from IRS 5, which when combined with radial measurements, can be
modeled as two distinct outflows aligned roughly north--south
\citep{IMAI:00, IDS:02}.  Infrared spectroscopy of CO fundamental
absorption features toward IRS 5 revealed three blueshifted components
suggestive of multiple outbursts \citep{MMH:91}.  From their 0.87 mm
line survey, \cite{HvD:97} found that W3 IRS 5 is a region in which
shock chemistry is present.

Relatively weak radio emission toward IRS 5 was first detected by Colley
(1980), who named the radio source W3~M.  More recent VLA
observations have resolved W3~M into a cluster of seven distinct
centimeter radio sources with fluxes around $\sim 1$~mJy, distributed over
a region 3$''$ (5000 AU) in diameter (TGC97, \citep[][hereafter
CG94]{CGJW:94}).

The nature of the radio source remains enigmatic.  These are examples of a
growing number of H II regions  with sizes less than few 10$^3$ AU, called
hypercompact HII regions (HCHII) \citep{GGDW:95}. An
outstanding question is the relationship between the two infrared sources
and the seven sources detected in the radio.  One possibility is that most
of the radio sources are too deeply embedded to be detected in the infrared. An
alternative is that the radio sources trace shock ionized regions in the
W3 IRS 5 outflow that are not luminous in the infrared. 
To further investigate the nature of the radio sources, and to better
understand the relationship between the radio and infrared sources, we
have obtained an additional epoch of VLA measurements of IRS 5.  The first
detailed images of the IRS 5 region were produced with the VLA A-array at
2~cm in 1989 January (CG94) multi-frequency, multi-array, VLA study carried out by
TGC97 with observations at 6, 2, and 1.3~cm from 1989 January to September.
The shortest wavelength observed by both CG94 and TGC97 was 1.3~cm. The
HCHII regions may be optically thick at 1.3~cm, therefore measurements at
0.7~cm, where dust emission should not contribute to the free--free
emission, are of great value in determining parameters such as emission
measure, electron density and the mass of ionized gas for these sources.

We have measured W3 IRS 5 with the VLA B-configuration at both 1.3 and
0.7~cm.  If the HCHII regions are ionized by embedded stars, these
observations allow us to estimate the Lyman continuum photon flux and
to classify the exciting stars.  In addition, we can compare our
results with those of CG94, to look for variations in source position
and flux density over time that may be indicative of a shock origin
for the radio sources. Additional comparisons are made with archival
near IR data taken with the NICMOS camera on board the Hubble Space
Telescope.

\section{OBSERVATIONS AND REDUCTION}

We observed W3 IRS 5 using the VLA, which is maintained and operated
by the National Radio Astronomy Observatory (NRAO)\footnote {The
  National Radio Astronomy Observatory is a facility of the National
  Science Foundation operated under cooperative agreement by
  Associated Universities, Inc.}.  The B-configuration observations
occurred over an 8 hr period beginning 2002 July 11 at 23:00
LST. High-frequency continuum measurements at 1.3 and 0.7~cm were
conducted in dual-polarization mode using four intermediate
frequencies (IFs), two in left circular polarization and two in right
circular polarization.  Two adjacent 50-MHz bands were recorded for a
total spanned bandwidth of 100 MHz centered on 22485.1 MHz for the
1.3~cm measurements and 43364.9 MHz for the 0.7~cm measurements.  In
order to reduce the effects of atmospheric phase fluctuations, we used
the technique of {\it fast-switching} \citep{CH:97} between the
target, W3 IRS 5, and a phase calibrator, 0228+673, that is 5.3$^{\rm
  o}$ away from W3 IRS 5.  We observed with a fast-switched cycle time
of 110~s, with 80~s spent on the target and 30~s on
the calibrator.  Cycles were repeated for 30 minutes at 0.7~cm
followed by 30 minutes at 1.3~cm.  A reference pointing scan at 3.5~cm
on 0228+673 was performed prior to the start of each 0.7~cm 30-minute
scan.  Finally, two scans were recorded each at 1.3 and 0.7~cm on the
calibrator source 0713+438, which is recommended by the NRAO for use
in determining the absolute flux density calibration at high
frequencies and is regularly monitored by the VLA at both 1.3 and
0.7~cm.
 
Data were reduced using the standard routines within the Astronomical
Image Processing System (AIPS).  The absolute flux density scale was
established using fluxes for 0713+438 determined by the VLA in
B-configuration on 2002 June 24, roughly two weeks prior to our
observations.  These flux densities were 0.506$\pm$0.002~Jy at 1.3~cm
and 0.263$\pm$0.003~Jy at 0.7~cm.  Target source phases were
calibrated by interpolating the phases from the fast-switched phase
calibrator source 0228+673.  From the amplitude and phase calibrated
data, a number of images of both the large scale structure of the W3
region and the compact sources in the W3 IRS 5 region were produced.
Figures \ref{KBAND_IMAG} and \ref{QBAND_IMAG} show the compact radio
emission toward W3 IRS 5 at 1.3~cm and 0.7~cm respectively.  For
purposes of comparison both images were produced using a circular
$0.2''$ beam.  Also plotted in Figures \ref{KBAND_IMAG} and
\ref{QBAND_IMAG} are crosses which represent positions of features
observed by CG94 and TGC97 along with the corresponding letter
designations.  Our contour maps show four distinct regions of emission
near the CG94 positions for sources IRS 5a, b, d1/d2 and e/f.

In order to extract relevant parameters such as source positions and
flux densities, two-dimensional Gaussian functions were fitted to the
four features visible in Figures \ref{KBAND_IMAG} and \ref{QBAND_IMAG}
using the AIPS task JMFIT.  In addition, we determined upper limits on
the flux densities corresponding to CG94/TGC97 sources IRS 5c, d1, e,
and g.  We accomplished this by integrating image flux densities over
a $0.2'' \times 0.2''$ box centered on the CG94/TGC97 positions using
the AIPS task IMEAN.  This task was complicated by blending in the
case of feature IRS 5d1 (discussed below).  The results from the fits
to the images analysis and the interpretation are discussed in the
next section.

\section{RESULTS AND DISCUSSION}

\subsection{Proper Motions}

In Table \ref{GAUS_FIT_TAB} we give the Gaussian fit results for the
four sources visible in our 1.3~cm and 0.7~cm data.  The most intense
source at both wavelengths is source IRS 5d2.  Since our images do not
completely resolve CG94 sources d1 and d2, it is possible that the
position and flux density for d2 in Table \ref{GAUS_FIT_TAB} suffer
from some blending with the weaker source d1. In the following we
refer to this as d1/d2.  Examination of Figures \ref{KBAND_IMAG} and
\ref{QBAND_IMAG} shows that the most northern source in our images
appears to be equidistant from the CG94 positions for IRS 5e and f.
Because CG94 found that IRS 5f had twice the flux density of IRS 5e at
2~cm, we believe it more likely that the stronger of the two sources
brightened while the weaker source is no longer seen.  We therefore
identify this peak with CG94 source IRS 5f in subsequent discussion.

In Figures \ref{KBAND_IMAG} and \ref{QBAND_IMAG} there appears to be a
slight ($\sim$0.02$''$) shift in $\alpha \cos(\delta)$ between two of
the peaks in our images and the CG94 sources d1/d2 and b.  Since we
have no absolute reference with which to register the two sets of
positions, we have chosen to compare relative angular distances
between features, which we term arclengths.  For example, the distance
between source d1/d2 and source f is arclength f-d1/d2. In Table
\ref{ARC_TAB} we compare the arclengths derived from our 1.3~cm and
0.7~cm data with those computed from the CG94 data at 2~cm.
Examination of the table shows that the arclength d1/d2--b remained
nearly unchanged to within 19~milli-arcseconds (mas) or less.
However, f-d1/d2 and f--b both show offsets greater than 200~mas at
comparable positions angles.  These differences imply that sources IRS
5d1/d2 and b were stationary between the epoch of our observations and
that of CG94, and that source IRS 5f is moving away from IRS 5b and
d1/d2 in a direction $\sim$12$^{\circ}$ east~of~north.  Over the time period
between the two epochs the $\sim$210~mas offset translates to a rate
of $\sim$15.6~mas~yr$^{-1}$.  The estimated uncertainty in the
arclength differences is 10\%. At the assumed distance to W3 IRS 5,
the proper motion is 135~km~s$^{-1}$.  The angular separation between
IRS 5f and d1/d2 in our 1.3~cm image is 0.56$''$.  The proper motion
of IRS 5f is in a direction approximately away from IRS 5d1/d2. If IRS
5f originated from IRS 5d1/d2, and the velocity of IRS 5f remained
constant, the travel time from IRS 5d1/d2 to the Epoch 2002.5 position
of IRS 5f is 36 years.

For source IRS 5a the position uncertainties are greater in Table
\ref{GAUS_FIT_TAB} because of the lower signal-to-noise ratios for the
peaks, especially at 1.3~cm.  Nevertheless, the measured position of
IRS 5a using the 1.3 cm data and that using the 0.7 cm data are in
good agreement, as seen in Table \ref{ARC_TAB}.  Both show a proper
motion of IRS 5a relative to stationary IRS 5b and IRS 5d1/d2.
Offsets between our 0.7~cm positions and those of CG94 are
$\sim$190~mas at a position angle of 50$^{\circ}$ E of N.  This offset
translates to a motion of $\sim$14.1~mas~yr$^{-1}$ or 122~km~s$^{-1}$
at the assumed distance to W3 IRS 5.  The magnitude of this velocity
is similar to that of component IRS 5f.

Since there may be additional motion along the line of sight, the
measured velocities of IRS 5f and IRS 5a are lower limits.  Given the
large magnitude of the velocities, it is unlikely that IRS 5f and IRS
5a contain stars. However, such velocities are often found in
outflows. Thus we propose that these components formed in outflows
originating in the W3 IRS 5 system.  As surveyed in the Introduction,
there is ample evidence that W3 IRS 5 is driving an outflow.  The
evidence includes the presence of a CO outflow and an expanding
cluster of H$_2$O masers (\cite{IMAI:00}).  \cite{IMAI:00} studied the
proper motions of the H$_2$O masers toward the IRS 5 region and
determined proper motions for 108 maser features some of which are
near the continuum sources IRS 5a, b, c and e. They concluded that
these motions are caused by two distinct outflows aligned north--south
with origins approximately 700~mas south of IRS 5d1/d2 and 300~mas
north of IRS 5a.  Since the motion of source IRS 5f is in agreement
with this north--south direction, this lends support to our
identification of our northern source with CG94 source IRS 5f.  The
alternative would be motion of source IRS 5e in a direction counter to
the outflow directions suggested by the maser proper motions.  We
consider this scenario unlikely.  Near source IRS 5a, there is a group
of H$_2$O masers with proper motions directed to the west, although
the speeds implied by the H$_2$O maser proper motions are $\sim$20\%
of the speed inferred for IRS 5a.  Our data show that source IRS 5a is
moving in nearly the opposite direction from these nearby masers with
a proper motion $\sim$50$^{\circ}$ east~of~north.  This position angle
is actually closer to the CO outflow direction of 38$^{\circ}$
observed by \cite{MMH:91} than to the two suggested north--south maser
outflows. We show a superposition of the locations of IRS 5d1/d2, b, a
and f, together with the H$_2$O masers and near IR sources in Fig.~3.

It is interesting that the two continuum sources for which we observe
proper motions have relatively few H$_2$O masers in their vicinity and
that these few masers have motions that are peculiar to the general
north--south flow.  In the case of source IRS 5f, the closest maser is
a single feature near CG94 component IRS 5e with a motion toward the
SE.  Thus we find an indication of an anti-correlation of continuum
source proper motions and presence of water masers. \cite{IMAI:00}
suggest that the maser clusters apparent in W3 IRS 5 (see Fig.~3) are
formed in a compressed shell of turbulent post-shock gas swept up by
outflows. We find it plausible that the continuum sources IRS 5a and f
are part of supersonic outflows sweeping up the ambient gas.

\subsection{Source Flux Densities} 

In Table \ref{FLUX_TAB} we compare integrated source flux densities
from our data and that of CG94 and TGC97. For sources IRS 5a, b, d2,
and f, our reported flux densities are the results of the Gaussian
fits to the peaks.  For sources IRS 5c, d1, e and g we used the AIPS
task IMEAN as discussed in Section 2.  For sources IRS 5c, e, and g
the IMEAN results were below the 3$\sigma$ rms noise level in the 1.3
and 0.7~cm images and thus represent upper limits on the flux
densities.  As mentioned in Section 2, the IMEAN result for
source IRS 5d1 should be viewed as an upper limit since there is
blending from source IRS 5d2. Also, the flux density of d2 may contain
a contribution from d1. The values in Table \ref{FLUX_TAB} longward of
2~cm are taken from TGC97.  For IRS 5g, all of the flux densities are
compiled by TGC97.  CG94 have tabulated flux densities for individual
sources at 2~cm but their 1.3~cm images did not resolve all features.
Instead, they cataloged 1.3 cm flux densities for three sources
designated a+b, c+d1/d2 and e+f.  Summed flux densities for these
sources are also reported in Table \ref{FLUX_TAB}. At 1.3~cm, we make
comparisons on a source-by-source basis with the data of TGC97. Our
flux densities for sources IRS 5a, b, d2 and f are a factor of 2--4
larger than those of TGC97, while the flux density of IRS 5d1 is
approximately equal to or lower than that of TGC97.  For sources IRS
5e and g, which we did not detect, our 3$\sigma$ upper limits are 0.6
and 0.8 times the values determined by TGC97.

It is unlikely that random errors are the cause of the differences,
given the signal to noise ratios. Recent improvements in VLA
instrumentation (i.e. K-band receiver upgrades) and data taking
techniques (i.e. fast-switching) provide greatly improved sensitivity
and reduced atmospheric decorrelation for data taken at shorter
wavelengths.  Thus it is possible that for IRS 5a, b, d2 and f, one
might expect our measured flux densities to be higher, even if the
sources were not varying over time.  However, sources IRS 5e and g,
which were observed by TGC97 but are undetected in our 2002
observations, provide evidence for time variability of the continuum
sources in the W3 IRS 5 region.

At longer wavelengths,  the TGC97 data show that  for sources IRS 5d1,
d2,  e and f  the 6~cm  flux densities  are greater,  on average  by a
factor  of 2,  than those  at 2~cm.   This may  be an  indication that
synchrotron  emission  contributes to  the  radio  emission at  longer
wavelengths.   Synchrotron  emission   has  been   found   toward  the
Turner-Welch    source   in   the    W3(OH)   star    forming   region
(\cite{RAMMM:95}).  Presumably  high   magnetic  fields  enhance  this
emission.  \citep{WRM:99}  have carried out  detailed, high resolution
measurements of the TW object in the W3(OH) region at 8.4 GHz with the
VLA.  There  are some similarites with  IRS 5, since the  TW source is
driving  an  H$_2$O  maser  outflow  and has  thermal  dust  emission.
\cite{FTCCK:93} observed a  population of HCHII regions  in the Orion
Nebula   with  non-thermal   spectral  indicies   and   variable  flux
densities.  They   speculated  that  these  changes   were  caused  by
photospheric  phenomena.   It is  possible  that  we  are measuring  a
similar type of  phenomenon, although the flux densities  of the IRS 5
sources are  about 20 times larger  than those in Orion,  and from our
Gaussian fits, the IRS 5 sources are extended.

>From our integrated flux densities at 1.3 and 0.7~cm, we have
calculated spectral indices for sources IRS 5a, b, d1/d2, and f.
Given the degree of blending, we now refer to d1 and d2 as
d1/d2. Since the measurements were almost simultaneous, these results
are not affected by variability.  For IRS 5a, we find a negative
index, which may indicate nonthermal emission. For the remaining three
sources we derived positive spectral indices between 1.3 and 0.7~cm.

To calculate source parameters for sources IRS 5a, b, d1/d2, and f, we
used our 1.3~cm data given in Table \ref{GAUS_FIT_TAB} and
relationships in the Appendix.  The results are presented in Table
\ref{PARAM_TAB}.  Although it is likely that some flux density
variations have occurred, the data in Table \ref{PARAM_TAB} give us
fairly reliable estimates.  For these calculations, we have assumed an
electron temperature of 10$^{4}$ K and uniform-density spherical
sources. For deconvolved sizes less than half of the FWHP beam, we used
$0.5 \times$  the beamsize.  Sources in our data are barely resolved, so the
formulae in \cite{MH:67} were replaced by those in \cite{PW:78}.  The
use of the relations of \cite{PW:78} means that we have radii that are
twice the values that one obtains by simply setting source diameters
equal to deconvolved FWHPs.  The emission measures are very large, and
the masses of ionized gas small.  This would indicate that all of
these are very young objects. We have not assigned stars for IRS 5a
and f since these sources have clearly changed position since 1989.

Given the lack of proper motion, and the coincidence with IR sources
(see Fig.~3), we find it likely that IRS 5d1/d2 and IRS 5b are ionized
by internal OB stars. We classify the exciting stars of these regions
by assuming that the Lyman continuum photons are not absorbed by dust
and that the 1.3 cm emission is optically thin. If so, these regions
are excited by early B stars. However the positive spectral indicies
of the IRS 5d1/d2 and IRS 5b regions suggest that at least a part of
the emission from these HCHII regions is optically thick. The
strongest case can be made for IRS 5d1/d2.  If the 1.3 cm flux density
is due to optically thick free-free emission, but the 0.7 cm emission
is optically thin, the Emission Measure will be a factor of 4, and an
rms electron density a factor of 2 larger than the values in Table
4. It is possible that at 0.7 cm this nebula may still have a large
free-free optical depth; such appears to be the case for the HCHII
regions in W49 which show a continuous rising power law spectrum from
1.3 cm to 3 mm (\cite{DWGW:01}). If IRS 5d1/d2 has optically thick
free-free emission even to 3 mm and we extrapolate the 3 mm flux
density using the measured spectral index, then the Emission Measure
and rms electron density could be a factor of 15 and 4, respectively,
larger than the values in Table 4.

IRS 5d1/d2 is the source with the largest increase in flux density,
2.3 mJy, between 1.3 and 0.7~cm (see Table \ref{GAUS_FIT_TAB}).  It is
unlikely that the increase in flux density between 1.3 and 0.7 cm is
caused by thermal dust radiation for several reasons.  First, the
spectral index is too small for dust emission, although this could be
a mixture of free-free and dust radiation. Second, from equation~7 in
the Appendix, we can determine the column density of protons, N(H). In
a 0.2$''$ beam, for a flux density of 3 mJy and dust temperature of
500 K, we have N(H)=$5\times 10^{26}$~cm$^{-2}$.  This is an
extraordinary value; this column density must give rise to an
extinction of greater than 10$^4$ visual magnitudes, so one needs a
very special geometry to detect IRS 5d1/d2 even in the infrared.  In
the same fashion, assuming that the smaller increase in flux density
found for IRS 5b is caused by dust emission and applying the same
analysis, we find that the resulting extinction would be greater than
$10^3$ magnitudes. This is also an extremely large value; these HII
regions are coincident with the 2$\mu$m. Hence it is unlikely that the
positive spectral indicies determined for these HII regions are caused
by thermal dust emission.

\subsection{Comparison with Near IR Data for W3}

We have overlayed the positions of our VLA sources on a NICMOS 2.22
$\mu$ meter image from the {\it HST} archive in Figure 3.  This image
was produced using the CALNICA V3.1.1 pipeline.  The data, taken in
the F222M filter, clearly resolve the infrared double source first
reported by \cite{HML:81}.  For IRS 5d1/d2 and IRS 5b, we find an
excellent agreement between the R.A. and Declination arclengths
measured from the NICMOS data and the VLA data.  The two pairs of
arclengths differ by only 0.007$''$ and 0.002$''$ in R.A. and
Declination, respectively.  These differences are within the errors of
the VLA positions.

The
difference in the absolute positions derived from the VLA and NICMOS
data is 0.4$''$ in Declination, which is within the expected
uncertainties in the absolute pointing of NICMOS.  After
refining the absolute pointing of the NICMOS data using this offset,
we find  no  sources at the positions of IRS 5a
or IRS 5f in the NICMOS image.
To corroborate the refinement of the NICMOS astrometry using the VLA
data, we compared the refined NICMOS coordinates for an optically
visible star with coordinates listed in the USNO A2 and B 
catalogs. We find offsets (R.A., Dec.) between our NICMOS position 
and
those of the USNO A2 and B catalogs of (-0.20$''$, 0.03$''$) and
(0.40$''$,  -0.22$''$) respectively.  The USNO positions from the two 
catalogs differ by (-0.60$''$, 0.25$''$), bracketing the NICMOS
positions.  Hence, we find a reasonable correspondence between the
refined NICMOS positions and USNO catalog positions for this star.
Given the excellent agreement between the relative offsets in the VLA
and NICMOS data, we conclude that the near-IR double is spatially
coincident with VLA sources IRS 5d1/d2 and IRS 5b.

\subsection{The Nature of IRS 5d1/d2 and IRS 5b}

These data provide a new insight into the nature of the radio sources
and the relationship between the radio sources and infrared
sources. The lack of proper motion between the radio sources IRS
5d1/d2 and IRS 5b, and the spatial coincidence found between the radio
and near-IR positions shows that these trace ionized gas close to
young OB stars.  In contrast, the sources IRS 5a and f show
significant proper motions in different directions, suggesting that
these trace ionized gas in outflows driven by the embedded O--B stars
ionizing IRS 5d1/d2 and IRS 5b.

The deconvolved angular radii of the IRS 5d1/d2 and IRS 5b radio
sources are less than 240~AU.  This is a generous limit based on
spherical source geometry. It is only 1.3 times the radius at which
the escape velocity for a 10M$_\odot$ star is equal to 10 km s$^{-1}$,
the sound velocity of the ionized gas.  Thus it is possible that the
ionized gas in IRS 5d1/d2 and IRS 5b is gravitationally bound to their
exciting stars.  The radius at which the escape velocity equals the
thermal velocity is a critical point for both accretion flows in HII
regions \citep{KETO:02} and stellar winds \citep{PARKER:58}; at this
radius the wind/accretion flow transitions from subsonic to supersonic
velocities.  It is possible that the HII region can be trapped by
accretion when the size of the Str\"omgren sphere is within this
critical radius \citep{KETO:02}.  In both the accretion and wind
models, the density of ionized gas rises toward the star, resulting in
strongly peaked radio emission.  The small size scales may also be
indicative of photoevaporating disks \citep{HJLS:94}, in which the
size of the optically thick HII region is on the order of 200 AU at
1.3 cm for an O7 star, assuming a weak stellar wind.

It is impossible at the moment to distinguish between these models.  A
feature of each of these models is a decreasing ionized gas density
with radius, which can result in an increase of flux density with
frequency (i.e. positive spectral indices) if the cores of the HII
regions are optically thick at the observed frequencies.  Hence, these
models can qualitatively explain the increase in flux seen between the
1.3 and 0.7 cm flux densities, although detailed modeling is needed to
explore the stellar and gas properties required to reproduce the
observed flux densities.  A large optical depth is attractive given
the mismatch between the high total infrared luminosity of IRS 5
\citep[$10^5$ L$_\odot$;][]{CBHECS:95} indicating the presence of an
O7 or O6 star, and the ionized gas parameters from the 1.3 cm data, which
are consistent with B1-B0 stars.  In this case, IRS 5d1/d2 with its
large positive spectral index may be the more massive of the two
sources, which is consistent with this source having both higher radio
and 2~$\mu$m flux densities than IRS 5b.

Although we cannot currently distinguish between wind and accretion
models, the small size of these regions does demonstrate that models
of hypercompact HII region such as IRS 5d1/d2 and IRS 5b must include the
influence of gravity from the exciting stars.  In this way, models of
hypercompact regions may differ significantly from models of
ultracompact HII regions.

It is likely that the system is currently bound by the combined mass
of the molecular gas and the two young stars, although it is not clear
whether these will remain bound once the gas is dispersed. At the
projected separation, the orbital velocity of a bound system would be
$3-5$ km s$^{-1}$, assuming masses of 10 M$_\odot$ for IRS 5b and
10--50 M$_\odot$ for IRS 5d1/d2.  In comparison, the full width at
half maximum velocity measured in NH$_3$ observations with a
$\sim3''$~beam, which provides a reasonable upper limit to the
relative velocities of the embedded stars, is 6 kms$^{-1}$
(\citep{TGW:98}). This is slightly larger than the required orbital
velocity.

\subsection{Nature of IRS 5a and IRS 5f}

Perhaps the best studied example of radio continuum sources in an
outflow are the radio jets associated with the HH 80-81/GGD 27 complex
(\cite{MRR:98}, \cite{MRR:95}). The driving source is IRAS 18162-2048,
with a luminosity of $2 \times 10^4$~L$_\odot$.  Symmetrically
distributed in a line about IRAS 18162-2048 are seven radio continuum
knots.  >From proper motions, these have velocities ranging from 500
to 1300 km s$^{-1}$, away from IRAS 18162-2048.  There is a smooth
decay of the free-free emission intensity with time for the two
innermost components.  In contrast, toward W3 IRS 5 we find no
evidence for the symmetric, ordered motions apparent in HH 80-81.
Although d1/d2 lies between the two radio sources with proper motions,
IRS 5f and IRS 5a, source IRS 5a shows proper motions of a similar
magnitude but in a roughly similar direction. If IRS 5d1/d2 is the
{\it only} source driving the outflow, the motion of IRS 5a is
difficult to explain, since one would expect source IRS 5a to be
moving in the {\it opposite} direction.  It is possible that there is
a second outflow source, which may be IRS 5b.  \cite{IMAI:00} had
reported two outflow centers.  However, our second outflow center must
be south of the position of IRS 5a, so this center must have a
different location from the second water maser outflow source, which
is north of IRS 5a.  See our superposition of radio continuum sources,
H$_2$O masers and near IR sources in Fig.~3. The lack of proper motion
of continuum sources south of IRS 5d1/d2 and IRS 5b may result from an
asymmetry in the surrounding dense molecular gas. There is also a
distinct absence of masers directly south of IRS 5b (\cite{IMAI:00}),
which could also be explained by a lack of dense gas.

IRS 5a and f may be the result of either high-speed outflows,
impinging on neutral dense clouds or ionized material ejected from the
outflow sources.  For electron densities of 10$^5$~cm$^{-3}$, the
recombination time is on the order of years, and is shorter than the
36 year travel time estimated for IRS 5f. However, \cite{MRR:98} have
shown that the moving sources in the HH~80-81/GGD~27 complex can be
fitted with a model in which the free-free emission decays on a long
timescale, and it is plausable that IRS 5a and IRS 5f are ionized
globules undergoing a similar decay.  \cite{DM:93} have reviewed
interstellar shock phenomena.  \cite{DS:96} and \cite{GH:98} have
carried out simulations of fast shocks. The results show that a
hundred km s$^{-1}$ outflow can ionize neutral hydrogen.  It is likely
that the geometry of the outflows in W3 IRS 5 are conical regions in a
northeast--southwest direction.  The interior of this cone would be
filled with low density gas. One can surmise that the CO outflow and
H$_2$O masers are located toward the edges of this cone.  More
energetic phenomena involving ionized gas, such as IRS 5f and IRS 5a
are found closer to the axis of cone, where the velocities are
larger. In support of this picture, we find that the proper motions of
the radio continuum sources are more than a factor of 2 larger than
those measured for the H$_2$O masers. The molecular material studied
by \cite{HvD:97} is easily dissociated, so must be farther from the
high speed outflow. This may be located in disks associated with but
perpendicular to the outflows.

\section{CONCLUSIONS}

On the basis of the analysis of our observations of the W3 IRS 5
hypercompact H~II (HCHII) regions at 1.3 and 0.7~cm and comparisons
with previous data from CG94 and TGC97, we conclude the following:

\begin{enumerate}

\item{W3 IRS 5 sources d1/d2 and b exhibited no motion relative to one
  another to within $\sim$20~mas over the time period 1989.1 to
  2002.5.  We use these as a reference to investigate possible motions
  of sources IRS 5a and f.}

\item{The HCHII region IRS 5f is offset from the CG94 1989.1 position
  by $\sim$210~mas at a position angle of 12$^{\circ}$
  east~of~north. This translates to a proper motion of
  15.6~mas~yr$^{-1}$, or a speed of 135 km s$^{-1}$. }

\item{IRS 5a is offset from the CG94 1989.1 position by $\sim$190~mas
  at a position angle of 50$^{\circ}$ east~of~north.  This translates
  to a proper motion of 14.1~mas~yr$^{-1}$. At the assumed distance to
  W3 IRS 5, this is an apparent speed of 122 km s$^{-1}$.}

\item{Using archival NICMOS data, we identify IRS 5d1/d2 and IRS 5b
  with the near IR sources first found by \cite{HML:81}. Given their
  lack of proper motion and coincidence with near IR sources, we
  conclude that these HII regions contain embedded stars. Our VLA
  measurements give upper limits of 240 AU to the sizes of the HII
  regions. These sources may be ionized by early B or O stars; we have
  only lower limits to the Lyman continuum flux from our continuum
  data.}

\item{Given the high velocities of the IRS 5a and IRS-f regions and
  the lack of any infrared sources toward these regions, we propose
  that these two regions are formed in the outflows originating in the
  W3 IRS 5 system.  These two HII regions may consist of shock ionized
  gas, or may be ionized knots expelled from the massive stars in W3
  IRS 5.}

\item{The proper motions of the IRS 5a and IRS 5f suggest the presence
  of two outflows.  IRS 5d1/d2 may be the outflow source causing the
  proper motion of IRS 5f, while IRS 5b may be the outflow source
  responsible for IRS 5a.  Two different driving sources were also
  inferred from H$_2$O maser proper motions measured \cite{IMAI:00}.
  However, a fit to the proper motions of the H$_2$O masers results in
  an outflow source that is 300~mas north of IRS 5a, which is
  inconsistent with the direction of our proper motion for IRS 5a.}

\item{Both of the sources containing stars, IRS 5d1/d2 and IRS 5b have
  positive spectral indices.  It is unlikely that this is due to dust
  emission, and is likely the result of a centrally condensed HII
  region that is optically thick toward the center of the region.  A
  centrally condensed HII region could result from a wind or accretion
  flow.}

\item{Of the two sources with significant proper motions, IRS 5f has a
  slightly positive spectral index, which is significantly different
  from optically thin free-free emission, while IRS 5a shows a
  negative spectral index.  The negative spectral index may indicate a
  contribution of synchroton emission.}

\item{From previous data, the flux densities of the HCHII regions
IRS 5d2, d1, e, f, and g  are larger at 6~cm than at 2~cm. This is further
evidence that at longer wavelengths there may be a contribution to the
emission from synchrotron radiation.}

\item{There is significant evidence for time variability for the W3
  IRS 5 continuum sources between 1989.1 and 2002.5; the case is
  strongest for sources IRS 5e and g.}

\end{enumerate}

\section{Acknowledgements}
We thank M. Corbin(STScI) for help with the reduction of the NICMOS data from 
the HST archive. 

\section{APPENDIX \label{APPENDIX}}

To obtain the correct values for emission measure, EM, the temperature
T must be corrected for beam dilution. This is related to the main
beam temperature, ${\rm T_{MB}}=60 \times $S(mJy), by

\begin{equation}
     {\rm T= T_{MB} \, \left(\frac{\theta(observed)}{\theta(source)}\right)^2 }
\end{equation}

where ${\rm \theta(observed)}$ is the apparent source size, in Table
1, and ${\rm \theta(source)}$ is the deconvolved size. To obtain the
deconvolved source size, we used the geometric mean source size, and
the usual deconvolution formula,

\begin{equation}
{\rm \theta(source)^2=\theta(observed)^2-\theta(beam)^2}
\end{equation}

$\theta({\rm source})$ is the Full Width Half Power (FWHP) deconvolved
source size in arc seconds. From \cite{PW:78}, for sources smaller
than the beam, the conversion from Gaussian source FWHP to a spherical
source radius is $\theta({\rm spherical \, source}) = 0.9 \times \,
\theta({\rm source})$ To convert a size in arc seconds to a linear
radius in parsecs for a distance of 1.83 kpc, we used ${\rm
  R(spherical \, source)=0.008 \times \theta(spherical \, source)}$.
For free-free radiation from ionized gas,
\begin{equation}
     {\rm  T= T_e \, \left(1 \, - e^{\tau_\nu} \right). }
\end{equation}

Assuming that $\tau\ll1$, one can use the following expression for
${\tau_\nu}$ to obtain the Emission Measure, EM, and RMS electron
density, n$_{\rm e}$

\begin{equation}
{\rm  \tau_\nu = 8.235 \times 10^{-2} \left(\frac{T_e}{K}\right)^{-1.35}
                \left( \frac{\nu}{GHz} \right)^{-2.1}
                \left( \frac{EM}{pc \,~cm^{-6}} \right) \, a (\nu, T). }
\end{equation}

For ${\rm T_e}$=10$^4$ and $\nu$=22.485 GHz, we have ${\rm EM=2.13
  \times 10^5 \, T}$.  The correction ${\rm \alpha(\nu, T)}$ is
usually $\cong 1$.  If $\tau_\nu$ is small compared to unity, then the
rms electron density, ${\rm n_e}$ is given by:

\begin{equation}
{\rm  n_e=2.46 \, \left( \frac{T}{K}\right)^{0.5}\, \left(\frac{T_e}{K}\right)^{0.175} 
\left(\frac{\nu}{GHz}\right)^{1.05}
\left(\frac{R}{pc}\right)^{-0.5}} .   
\end{equation}

For ${\rm T_e}$=10$^4$ and $\nu$=22.485 GHz, we obtain
$${\rm n_e=324 \, \left(\frac{T}{K}\right) \,
  \left(\frac{R}{pc}\right)^{-0.5} }$$ To calculate the mass of
ionized gas we remove the extra electrons contributed by helium, whose
abundance is y$_{\rm n}$.  Then the mass is given by:

\begin{equation}
    {\rm  \left(\frac{M}{M_\odot}\right)=0.10 \, [1+y_m] \, \left( \frac{R}{pc} \right)^3 \, \left(\frac{n_e}{cm^3}\right) },
\end{equation}

where y$_{\rm m}$ is the mass fraction of helium; for y$_{\rm n}$=0.1,
y$_{\rm m}$=0.4. These calculations are for uniform density
spheres. If there are gradients or clumps, the value of ${\rm n_e}$ is
a lower limit and the mass is an upper limit to actual values.

For thermal emission from small dust particles, if the radiation is 
expressed in mJy, the source size, $\theta$ in arc seconds, peak flux density 
S$_\nu$ in mJy, and wavelength, $\lambda$ in mm, the column density of 
hydrogen in all forms, $N_{\rm H}$, in the Rayleigh-Jeans approximation, 
is given by the following relation \citep[see][]{RW:99}:
\begin{equation}
{\rm N_{H} = 1.93 \times 10^{24} \, \left(\frac{ S_\nu }{mJy}\right) \, \left(\frac{\theta}{''}\right)^{-2}
 \, \left(\frac{\lambda}{mm}\right)^4 \, \left(\frac{Z}{Z_\odot}\right)^{-1} \, \left(\frac{1}{b}\right) \,
\left(\frac{T_{dust}}{K}\right)^{-1}  \,.}
\end{equation}

In the centimeter and millimeter wavelength range, the dust optical
depth usually increases with $\lambda^{-2}$; then flux density
increases as $\lambda^{-4}$. If we take the metal abundance to be
solar, Z=Z$_\odot$, the dust properties appropriate for a very dense
gas, b=5 \cite[see][]{OH:94}, T$_{\rm dust}$=$5\times 10^2$~K,
$\lambda=7$mm and $\theta=0.2''$, we have $N(\rm H)=4.6 \times 10^{25}
\, S_\nu$

\clearpage

\begin{deluxetable}{lcccccccc}
\tabletypesize{\scriptsize}
\tablewidth{0pt}
\tablecaption{Gaussian fits to the  1.3 and  0.7~cm data. \label{GAUS_FIT_TAB}}
\tablehead{  & & & & \colhead{Major} & \colhead{Minor} & 
\colhead{Position\tablenotemark{a}    } & \multicolumn{2}{c}{Flux Density} \\
\colhead{W3 IRS 5}  & \colhead{$\lambda$} & \colhead{$\alpha$ (B1950)} & 
\colhead{$\delta$ (B1950)} & \colhead{Axis} & \colhead{Axis} & 
\colhead{Angle} & \colhead{Peak} & \colhead{Integrated} \\ 
\colhead{Source}  & \colhead{(cm)} & \colhead{2$^{\rm h}$ 21$^{\rm m}$} & 
\colhead{61$^{\circ}$ 52$'$} & \colhead{($''$)} & \colhead{($''$)} & 
\colhead{($^{\circ}$)}  & \colhead{(mJy/beam)} & \colhead{(mJy)} }
\startdata
a  \dotfill & 1.3 & 53.2196$\pm$0.0025$^{\rm s}$ & 20.939$\pm$0.036$''$ & 
0.399$\pm$0.084 &
0.196$\pm$0.041 &   1.30$\pm$11.01 & 0.654$\pm$0.014 &  1.279$\pm$0.036 \\
b  \dotfill & 1.3 & 53.2250$\pm$0.0009$^{\rm s}$ & 20.515$\pm$0.009$''$ & 
0.295$\pm$0.020 & 
0.208$\pm$0.014 &   1.11$\pm$7.65  & 2.055$\pm$0.014 &  3.158$\pm$0.033 \\
d1/d2 \dotfill & 1.3 & 53.3248$\pm$0.0003$^{\rm s}$ & 21.514$\pm$0.003$''$ & 
0.263$\pm$0.006 & 
0.193$\pm$0.004 &  11.66$\pm$3.01  & 6.169$\pm$0.014 &  7.816$\pm$0.029 \\
f  \dotfill & 1.3 & 53.4044$\pm$0.0009$^{\rm s}$ & 22.491$\pm$0.007$''$ & 
0.240$\pm$0.017 &
0.209$\pm$0.015 &   4.53$\pm$21.33 & 2.013$\pm$0.014 &  2.523$\pm$0.029 \\
& & & & & & & & \\
a  \dotfill & 0.7 & 53.2204$\pm$0.0016$^{\rm s}$ & 20.977$\pm$0.015$''$ & 
0.215$\pm$0.037 &
0.156$\pm$0.027 &  11.74$\pm$21.23 & 0.967$\pm$0.016 &  0.814$\pm$0.025 \\
b  \dotfill & 0.7 & 53.2236$\pm$0.0005$^{\rm s}$ & 20.492$\pm$0.005$''$ & 
0.239$\pm$0.012 &
0.187$\pm$0.009	&   0.23$\pm$7.84  & 3.398$\pm$0.016 &  3.804$\pm$0.031 \\
d1/d2 \dotfill & 0.7 & 53.3260$\pm$0.0002$^{\rm s}$ & 21.495$\pm$0.002$''$ & 
0.251$\pm$0.005 &
0.194$\pm$0.004	&  17.29$\pm$3.04  & 8.339$\pm$0.016 & 10.151$\pm$0.032 \\
f \dotfill & 0.7 & 53.4044$\pm$0.0010$^{\rm s}$ & 22.484$\pm$0.008$''$ & 
0.256$\pm$0.020 &
0.206$\pm$0.016	& 168.81$\pm$14.59 & 2.070$\pm$0.016 &  2.734$\pm$0.034 \\
\enddata
\tablenotetext{a}{ The uncertainty quoted was obtained from gaussian fits to the images. To obtain agreement with the maxima IRS 5d1/d2 and b from CGJW, a shift of $\sim$20 mas in $\alpha \cos(\delta)$ was needed. This may be the size of the systematic positional error. } 
\end{deluxetable}

\begin{deluxetable}{lccccc}
\tabletypesize{\footnotesize}
\tablewidth{0pt}
\tablecaption{Arclength differences between our data and CGJW. \label{ARC_TAB}}
\tablehead{  & \colhead{$\lambda$} & \colhead{$\Delta$ R.A.}  &  
\colhead{$\Delta$ Dec.}   & \colhead{ Total Offset} &  \colhead{Position 
Angle\tablenotemark{b}}\\
\colhead{Arclength\tablenotemark{a}}  & \colhead{cm} & \colhead{($''$)} & 
\colhead{($''$)} &  \colhead{($''$)}  & \colhead{($^{\circ}$)} }
\startdata
f-d1/d2  \dotfill & 0.7 &  0.045  &  0.209  &  0.214  &  12.2 \\
d1/d2--b  \dotfill & 0.7 &  0.002  &  0.013  &  0.013  &  10.2 \\
f--b   \dotfill & 0.7 &  0.047  &  0.221  &  0.227  &  12.1 \\
a-d1/d2  \dotfill & 0.7 &  0.123  &  0.143  &  0.188  &  40.7 \\
a--b   \dotfill & 0.7 &  0.125  &  0.155  &  0.200  &  38.9 \\
& & & & & \\
f--d2  \dotfill & 1.3 &  0.054  &  0.196  &  0.204  &  15.3 \\
d1/d2--b  \dotfill & 1.3 & -0.016  &  0.009  &  0.019  & 120.5 \\
f--b   \dotfill & 1.3 &  0.038  &  0.206  &  0.209  &  10.4 \\
a-d1/d2  \dotfill & 1.3 &  0.126  &  0.085  &  0.152  &  56.0 \\
a--b   \dotfill & 1.3 &  0.110  &  0.094  &  0.145  &  49.3 \\
\enddata
\tablenotetext{a}{Designates arclength between two respective W3 IRS 5 source 
components
for which we measured differences between our data and that of CGJW.}
\tablenotetext{b}{Position angle measured East of North.}
\end{deluxetable}

\begin{deluxetable}{lccccc}
\tabletypesize{\footnotesize}
\tablewidth{0pt}
\tablecaption{Comparison of W3 IRS 5 Source Flux Densities \label{FLUX_TAB}}
\tablehead{ \colhead{W3 IRS 5} & \multicolumn{4}{c}{Total Flux Density (mJy)} \\
\colhead{Source} & \colhead{6cm} & \colhead{2cm} & \colhead{1.3cm} & 
\colhead{0.7cm}}
\startdata
a  \dotfill & 0.6\tablenotemark{d} & 0.6\tablenotemark{d},~ 
0.52\tablenotemark{c} &
0.3\tablenotemark{d},~ 1.3\tablenotemark{a}    & 0.8\tablenotemark{a} \\
b  \dotfill & 0.8\tablenotemark{d} & 0.7\tablenotemark{d},~ 
0.78\tablenotemark{c} & 
0.6\tablenotemark{d},~ 3.2\tablenotemark{a}    & 3.8\tablenotemark{a} \\ 	
a+b \dotfill & \nodata  & \nodata  &  2.0\tablenotemark{c}  & \nodata \\
 & & & & & \\
c  \dotfill & 0.7\tablenotemark{d} & 0.6\tablenotemark{d},~ 
0.56\tablenotemark{c} & 
$<$0.3\tablenotemark{d},~ $<$0.45\tablenotemark{b}   &  $<$0.48\tablenotemark{b} \\
d1\dotfill & 1.7\tablenotemark{d} & 0.8\tablenotemark{d},~ 
0.48\tablenotemark{c} &
1.9\tablenotemark{d},~$<$1.5\tablenotemark{b}    & $<$2.1\tablenotemark{b} \\
d2 \dotfill & 2.6\tablenotemark{d} & 1.8\tablenotemark{d},~ 
1.02\tablenotemark{c} &
4.2\tablenotemark{d},~ 7.8\tablenotemark{a,e}    &  10.2\tablenotemark{a,e}   \\
c+d \dotfill & \nodata  & \nodata  & 2.5\tablenotemark{c}  & \nodata \\
 & & & & & \\
e  \dotfill & 2.0\tablenotemark{d} & 1.0\tablenotemark{d},~ 
0.36\tablenotemark{c} &	
0.7\tablenotemark{d},~ $<$0.45\tablenotemark{b}   &  $<$0.48\tablenotemark{b} \\
f  \dotfill & 1.8\tablenotemark{d} & 0.6\tablenotemark{d},~ 
0.72\tablenotemark{c} &
0.8\tablenotemark{d},~ 2.5\tablenotemark{a}    & 2.7\tablenotemark{a}     \\
e+f \dotfill & \nodata  & \nodata  & 1.4\tablenotemark{c}  & \nodata \\
 & & & & & \\
g  \dotfill & 0.5\tablenotemark{d} & 0.3\tablenotemark{d},~ 
\nodata\tablenotemark{c}  &
0.6\tablenotemark{d},~ $<$0.45\tablenotemark{b}   &  $<$0.48\tablenotemark{b} \\
\enddata
\tablenotetext{a}{From Gaussian fit to image peak.}
\tablenotetext{b}{Estimate from task IMEAN (see text).}
\tablenotetext{c}{Data from CGJW.}
\tablenotetext{d}{Data from TGCWJ.}
\tablenotetext{e}{Possible contamination from d1.}
\end{deluxetable}

\begin{deluxetable}{lcccrccccc}
\tabletypesize{\footnotesize}
\tablewidth{0pt}
\tablecaption{Parameters of the Ionized Regions \label{PARAM_TAB}}
\tablehead{  & \colhead{Spectral}&\colhead{Peak Flux} & \colhead{Angular\tablenotemark{a}} & 
\colhead{Peak\tablenotemark{b}} & \colhead{Emission} &     
& \colhead{Electron\tablenotemark{c}} &       &  \\
\colhead{W3 IRS 5} &\colhead{Index}& \colhead{Density} & \colhead{Size} & 
\colhead{Temp.} & \colhead{Measure} & \colhead{Radius} & 
\colhead{Density} &  \colhead{Mass\tablenotemark{c}}  &  
 \colhead{Spectral\tablenotemark{d}}   \\     
\colhead{Source} & &\colhead{(mJy/beam)} & \colhead{($''$)} & \colhead{(K)}  & 
\colhead{(cm$^{-6}$pc)} & \colhead{(pc)} & \colhead{(cm$^{-3}$)}  &
\colhead{(M$_{\odot}$)} & \colhead{Type} }
\startdata
 a  \dotfill & -0.69$\pm$0.06&0.6 & 0.28 & 80  & 1.7(7) & 1.6(-3) & 7.3(4) & 3.5(-5) & 
\nodata\tablenotemark{e} \\
 b  \dotfill & 0.28$\pm$0.02&2.0 & 0.25 & 340 & 7.2(7) & 1.2(-3) & 1.7(5) & 3.8(-5) & B0.5-
B1\\
d1/d2 \dotfill & 0.39$\pm$0.01 & 6.2 & 0.22 &1800 & 3.7(8) & 8.0(-4) & 4.8(5) & 3.1(-5) & B0.5-
B1\\
 f  \dotfill & 0.12$\pm$0.02&2.0 & 0.22 & 580 & 1.2(8) & 8.0(-4) & 2.8(5) & 1.8(-5) & 
\nodata\tablenotemark{e} \\
\enddata
\tablenotetext{a}{Geometric mean of angular sizes in Table \ref{GAUS_FIT_TAB}.}
\tablenotetext{b}{Corrected for beam dilution; if deconvolved size $<0.1''$, we  used 0.1$''$. see Appendix.}
\tablenotetext{c}{see Appendix.}
\tablenotetext{d}{see Rohlfs \& Wilson 1999}
\tablenotetext{e}{These are sources with proper motions; see text for details.}
\end{deluxetable}

\clearpage

\centering{Figure Captions}

\figcaption{Continuum image of the W3 IRS 5 region at 1.3~cm (Epoch 2002.5).  The FWHM beam 
size is $0.2'' \times 0.2''$. Contour levels are -1, 1, 2, 4 and 8 times 
the 3$\sigma$ RMS noise in the image of 0.45 mJy/beam.  The peak flux density
is 6.16 mJy/beam, which corresponds to a main beam brightness temperature of 
371.8 K.  The crosses designate positions of sources IRS 5a, b, c, d1, d2, e, 
and f from CGJW and the position of source IRS 5g from TGCWJ (Epoch 1989.1).  In our beam, IRS 5d1 
and d2 are blended. 
\label{KBAND_IMAG}}

\figcaption{Continuum image of the W3 IRS 5 region at 0.7~cm (Epoch 2002.5).  The FWHM beam 
size is $0.2'' \times 0.2''$. Contour levels are -1, 1, 2, 4, 8 and 16 times 
the 3$\sigma$ RMS noise in the image of 0.48 mJy/beam.  The peak flux density 
is 8.34~mJy/beam, which corresponds to a main beam brightness temperature of 
135 K.  The crosses designate positions of sources IRS 5a, b, c, d1, d2, e, and 
f 
from CGJW and the position of source IRS 5g from TGCWJ (Epoch 1989.1). In our beam, IRS 5d1 and 
d2 are blended. 
\label{QBAND_IMAG}}

\figcaption{NICMOS 2.22 $\mu$m image of the W3-IRS 5 region.
The greyscale image shows the near-IR double, with the positions of IRS 5d1/d2
and IRS 5b marked.  The orientation of the data is B1950,
the offsets are relative to the position of IRS 5d1/d2, 
A previously undetected source is apparent in the
NICMOS image between IRS 5d1/d2 and IRS 5b; this source appears extended.  The
positions of sources IRS 5a and IRS 5f are marked with arrows showing the
direction and 
magnitude of the proper motions between Epochs 1989.1 and 2002.5.  
Also 
shown are the positions of the H$_2$O masers \citep{IMAI:00}.  The lengths and
directions of the arrows show the expected proper motions of the masers
over a 100 year period relative to an arbitrary inertial reference center. Note
that the maser motions are shown over a 100 year period while the
continuum source motions are shown over a 13 year period.  The
proper motions of the continuum sources are typically a factor of 2  
more larger  than than those  of the masers. 
\label{NICMOS}} 

\clearpage

\begin{figure}[hbt]
\plotone{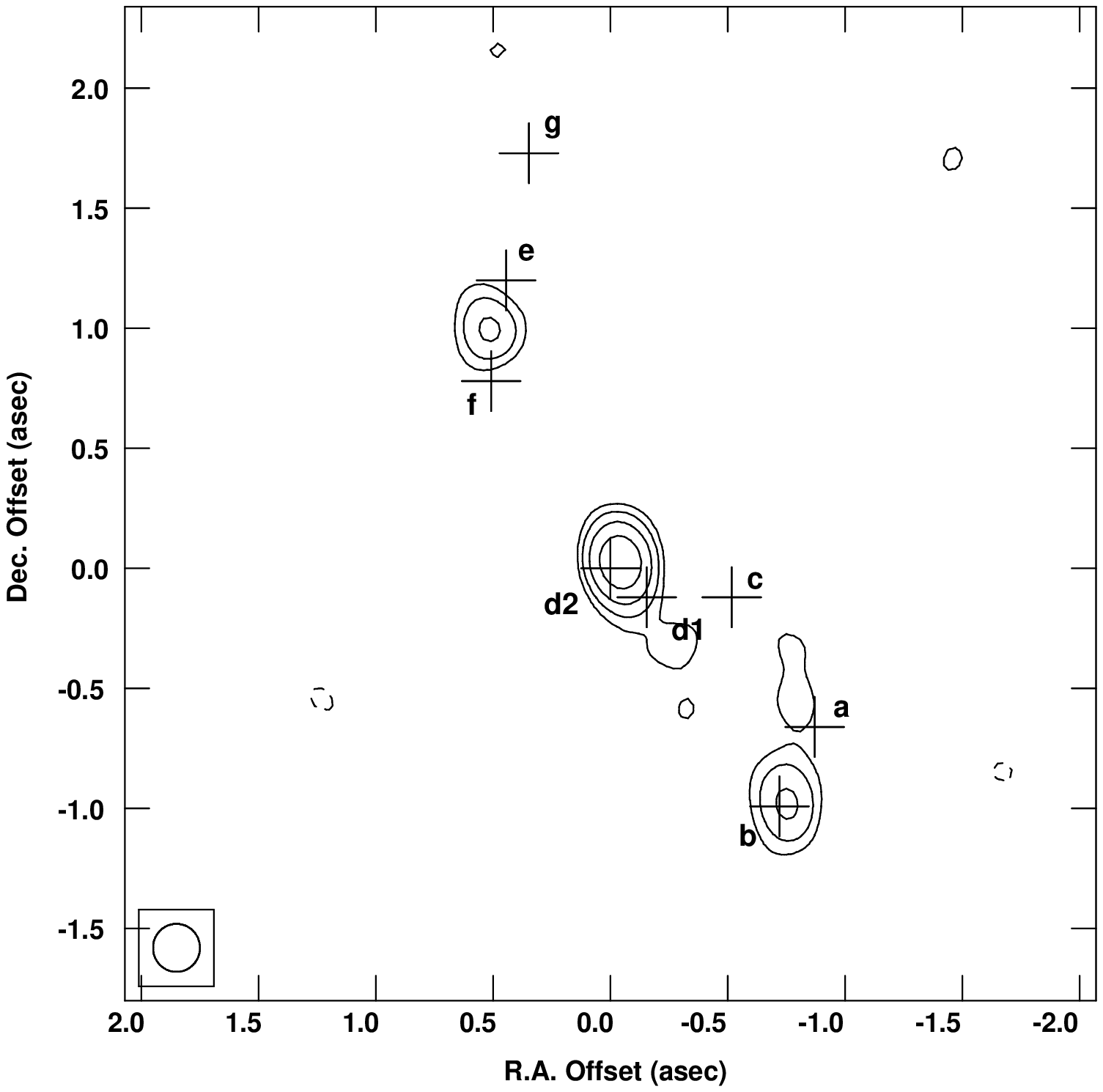}

\centerline{Figure 1}
\end{figure}

\begin{figure}[hbt]
\plotone{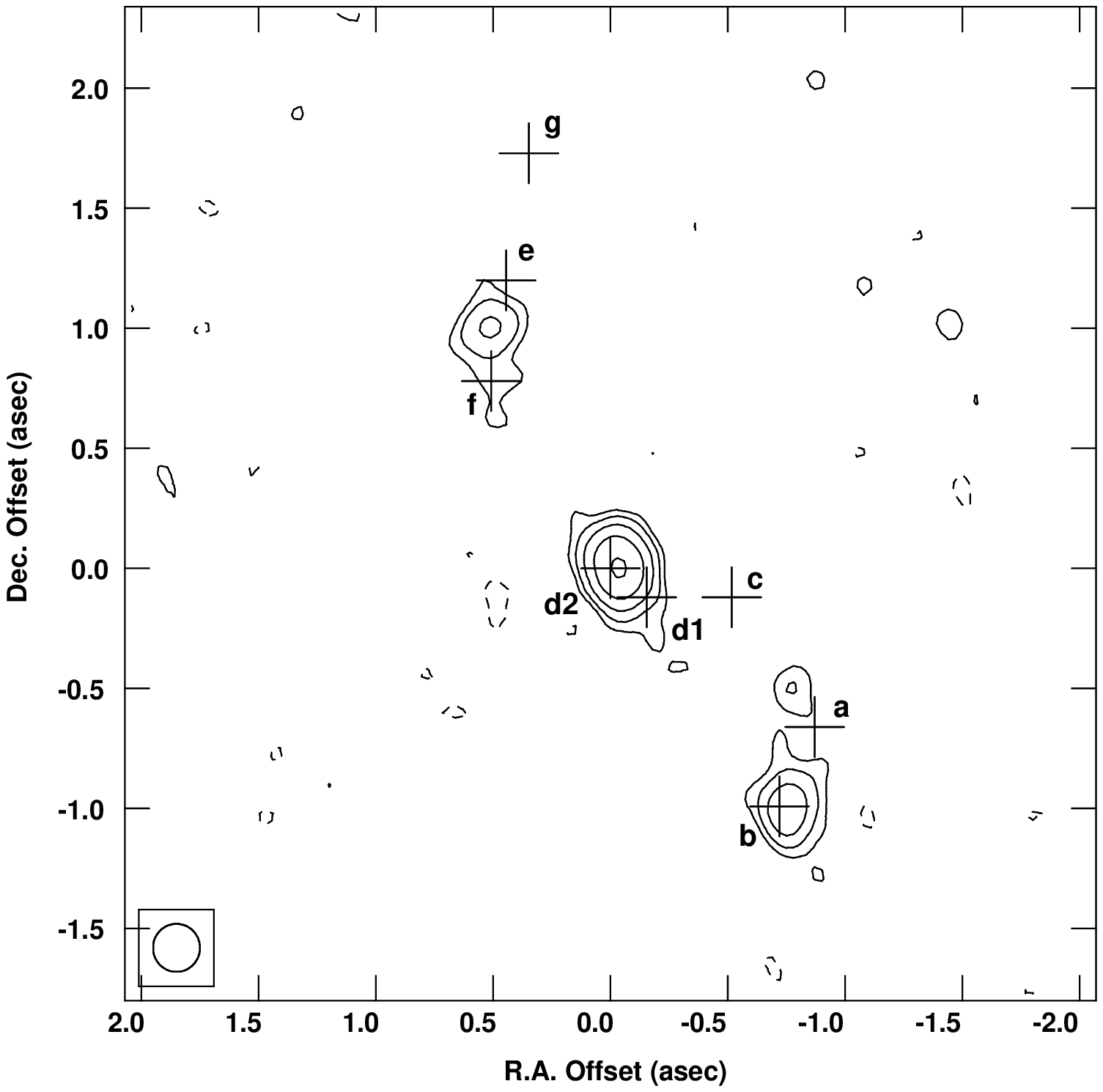}

\centerline{Figure 2}
\end{figure}

\begin{figure}[hbt]
\plotone{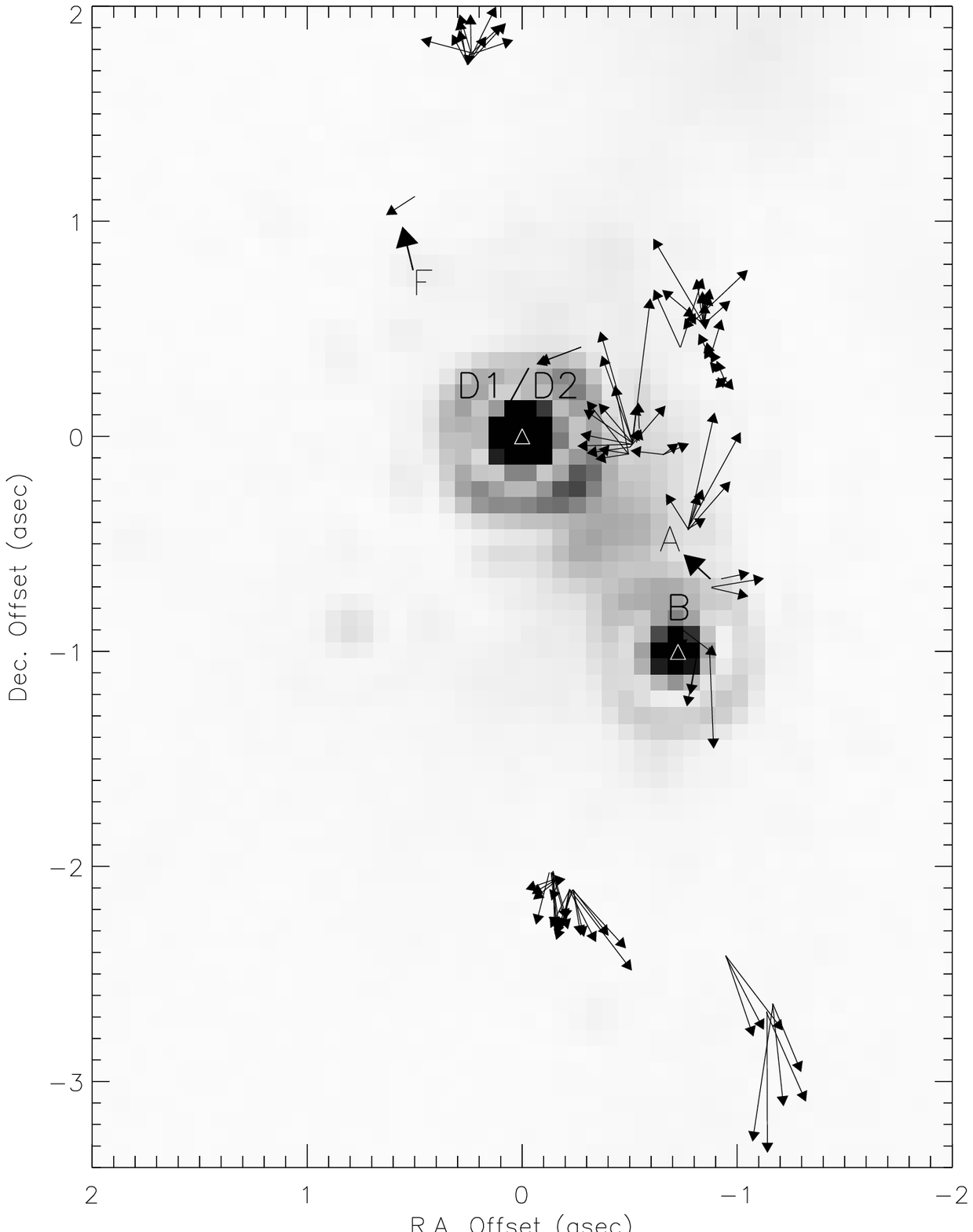}

\centerline{Figure 3}
\end{figure}

\end{document}